\documentclass[
    aps,
    fleqn,
    a4paper,
    superscriptaddress,
    preprint,
    preprintnumbers,
    showpacs,
    showkeys
]{revtex4}

\usepackage{graphicx}
\usepackage{picins}
\usepackage{amsmath}
\usepackage{amsfonts}
\usepackage[cp1250]{inputenc}

\newcommand{\bE}{\mbox{\boldmath$E$}}
\newcommand{\bP}{\mbox{\boldmath$P$}}
\newcommand{\bD}{\mbox{\boldmath$D$}}

\begin{document}

\preprint{{\em Ferroelectrics\/}: Vol. 223, No. 1-4, Pgs. 127-134 (1999) - {\em (Revised version)}}

\title{
Displacements of ${\bf 180^{\circ}}$ domain walls in electroded ferroelectric single crystals: \\
the effect of surface layers on restoring force
}

\author{A. \surname{Kopal}}
\email{antonin.kopal@tul.cz}
\affiliation{Dept. of Physics, International Center for Piezoelectric Research, Technical University Liberec, Liberec 1, 461 17 Czech Republic}
\author{P. \surname{Mokr\'{y}}}
\email{pavel.mokry@tul.cz}
\affiliation{Dept. of Physics, International Center for Piezoelectric Research, Technical University Liberec, Liberec 1, 461 17 Czech Republic}
\author{J. \surname{Fousek}}
\affiliation{Dept. of Physics, International Center for Piezoelectric Research, Technical University Liberec, Liberec 1, 461 17 Czech Republic}
\affiliation{Materials Research Laboratory, The Pennsylvania State University, State College, PA 16801, USA}
\author{T. \surname{Bahn\'{\i}k}}
\affiliation{Dept. of Physics, International Center for Piezoelectric Research, Technical University Liberec, Liberec 1, 461 17 Czech Republic}

\date{\today}

\begin{abstract}
Macroscopic properties of ferroelectric samples, including those in form
of thin films, are, to large extent, influenced by their domain
structure. In this paper the free energy is calculated for a plate-like
sample composed of nonferroelectric surface layers and ferroelectric
central part with antiparallel domains. The sample is provided with
electrodes with a defined potential difference. The effect of applied
field and its small changes on the resulting domain structure is
discussed. This makes it possible to determine the restoring force
acting on domain walls which codetermines dielectric and piezoelectric
properties of the sample. Calculations of the potential and free energy
take into account interactions of opposite surfaces and are applicable
also to thin films.
\end{abstract}


\maketitle

\section{Introduction}\label{artA1:sIntro}

It is known that samples of ferroelectric single crystals often
posses a surface layer whose properties differ from those of the
bulk. It may be a layer produced during the growth of a
crystalline plate or produced during the preparation of a
plate-like sample. Many observations gave evidence to the fact
that such a layer is either nonferroelectric or does not take part
in the switching process of the internal part; in any case its
permittivity is believed to differ from that of a homogeneous
sample in the ferroelectric phase. Its existence is expected to
greatly influence macroscopic properties of bulk
samples\cite{ArtA1:Miller,ArtA1:Drougard,ArtA1:Muser,ArtA1:Callaby} as well as
of thin films.\cite{ArtA1:Tagantsev} In this paper two such
consequences are investigated. First we reconsider the problem of
equilibrium domain structure in a ferroelectric sample possessing
a surface layer, previously discussed by Bjorkstam and
Oettel.\cite{ArtA1:Bjorkstam} Second, we evaluate the restoring force
acting on $180^{\circ}$ domain walls due to the layer; this will
make it possible to estimate the extrinsic contributions to
permittivity, piezoelectric coefficients and elastic compliances
of a ferroelectric sample. Investigations of crystals of the KDP
family revealed the existence of huge wall contributions to these
properties.\cite{ArtA1:Nakamura,ArtA1:Stula}

In previous papers on a related subject\cite{ArtA1:Drougard}, depolarizing 
field was considered as the source of energy which slows down the motion of a 
single domain wall in a dc electric field, as the wall departs from its 
original position by substantial distances. In contrast to such models we 
investigate very small deviations of walls forming a regular domain 
pattern.

\section{Geometry, variables and energy of the system} \label{artA1:sGeomEng}

{ 

We consider a plate-like electroded sample of infinite area with
major surfaces perpendicular to the ferroelectric axis $z$.
Central ferroelectric part with antiparallel domains (2.) is
separated from the electrodes (0.), (4.) by nonferroelectric
layers (1.), (3.) (see Fig.\,\ref{artA1:fig:geometry}). The
spatial distribution of the electric field $\bE$ is determined by
the applied potential difference $V=\varphi^{(4)}-\varphi^{(0)}$
and by the bound charge div$\bP_0$ on the boundary of
ferroelectric material, where $\bP_0$ stands for spontaneous
polarization. Geometrical, electrical and material parameters of
the system are shown in Fig.\,\ref{artA1:fig:geometry}.

We further introduce the symbols
\begin{displaymath}
    c = \sqrt{\varepsilon_a/\varepsilon_c}, \qquad
    g = \sqrt{\varepsilon_a\,\varepsilon_c},
    \nonumber 
\end{displaymath}
%
%
%
\parpic[r][t]{
    \begin{minipage}[t]{90mm}
    \begin{minipage}[t]{\textwidth}
        \makebox[\textwidth][t]{
            \includegraphics[width=85mm]{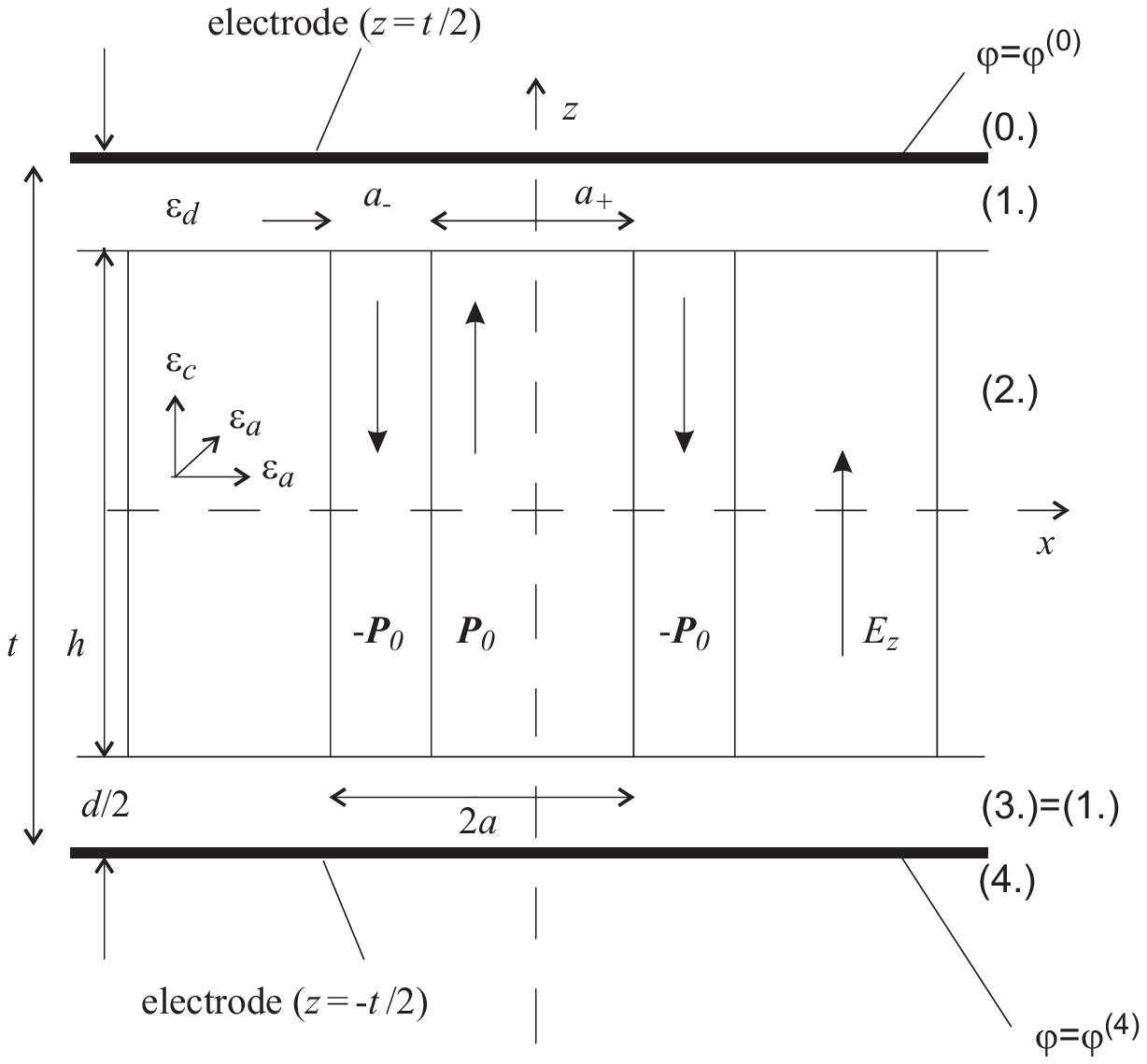}
        }
 \caption{Geometry of the model \label{artA1:fig:geometry}}
    \end{minipage}
    \end{minipage}
}
%
%
\begin{displaymath}
    S_0 = \sum_{n=1,3,5}^{\infty} \frac{1}{n^3} \doteq 1.052
\end{displaymath}
and several geometrical parameters: \label{artA1:GPar}
\[B=\frac{d}{h}\]
the domain pattern factor
\[R = \frac{\pi h}{2a}, \quad 2a = a_+ + a_- \]
and the asymmetry factor
\[A = \frac{a_+ - a_-}{a_+ + a_-}\ .\]

}  

The ferroelectric material itself is approximated by the equation
of state
\begin{eqnarray}
    D_{x} &=& \varepsilon_0\varepsilon_a E_{x},
    \nonumber \\
    D_z &=& \varepsilon_0\varepsilon_c E_z + P_{0},
    \nonumber
\end{eqnarray}
where $P_{0}$ is the spontaneous polarization along the
ferroelectric axis. This linear approximation limits the validity
of our calculations to the temperature region not very close below
the transition temperature $T_c$. Domain walls are assumed to have
surface energy density $\sigma_w$ and zero thickness.

\enlargethispage{1em}
The Gibbs electric energy of the system includes the domain wall
energy, the electrostatic energy whose density is $(1/2)\bE\cdot
(\bD-\bP_0)$ and the work performed by external electric sources
$-VQ$, where $Q$ is the charge on positive electrode. First,
Laplace equations have to be solved for electric potentials in the
bulk and in the surface layers, fulfilling the requirement of
potential continuity as well as conditions of continuity of normal
components of $\bD$ and tangential components of $\bE$. A rather
cumbersome calculation leads to the following formula for Gibbs
electric energy per unit area of the system (in $\rm J\,m^{-2}$):
\begin{equation}
    G =
        \frac{2}{\pi}\sigma_w R
        +
        P_0 A\
        \frac{
            P_0 A\, B t
            -
            2\varepsilon_0\varepsilon_d V \left( 1 + B \right)
        }{
            2\varepsilon_0
            \left(1 + B\right)
            \left(\varepsilon_d + B\,\varepsilon_c \right)
        }
        +
        \frac{4 P_0^2 t}{\varepsilon_0\pi^2 R (1+B)}
        \sum^{\infty}_{n=1}
        \frac{
            \sin^2{(n\pi(1+A)/2)}
        }{
            n^3(\coth{n B R} +g\coth{n R c})
        }\ .
    \label{artA1:eq:depolenergy}
\end{equation}

\eject

The first term represents domain wall contribution while the last one is
the depolarization energy. In the second term we recognize the effect of
layers (1.) and (3.) and of the applied voltage.

Let us compare expression with formulae deduced and used in
previous papers. For $V = 0$ and $s\to \infty$, $\varepsilon_d =
1$ and $A = 0$ the system goes over into an isolated ferroelectric
plate with ``neutral'' domain structure, placed in vacuum. In this
case the equation (\ref{artA1:eq:depolenergy}) reduces to the
expression given by Kopal et al.\cite{ArtA1:Kopal} for
ferroelectric plates of finite thickness in which the interaction
of the two surfaces is accounted for. If the plate is thick this
interaction can be neglected and Eq.\,(\ref{artA1:eq:depolenergy})
simplifies to the classical formula of Mitsui and
Furuichi\cite{ArtA1:Mitsui} (cf. Eq.\,(9) in Ref.\,
\cite{ArtA1:Kopal}) which is often used to determine the value
$\sigma_w$ from the observed width of domain patterns. Finally, in
the limit of $V = 0$ our formula (\ref{artA1:eq:depolenergy})
should converge to the expression deduced by Bjorkstam and
Oettel.\cite{ArtA1:Bjorkstam} In fact this is not the case and it
appears that the electric displacement as expressed in
Ref.\,\cite{ArtA1:Bjorkstam} does not satisfy all boundary conditions.

\section{Equilibrium domain structure for $V=0$}\label{artA1:sEDS}

\begin{figure}[t]
\begin{center}
\begin{minipage}{11cm}
 \includegraphics[width=11cm]{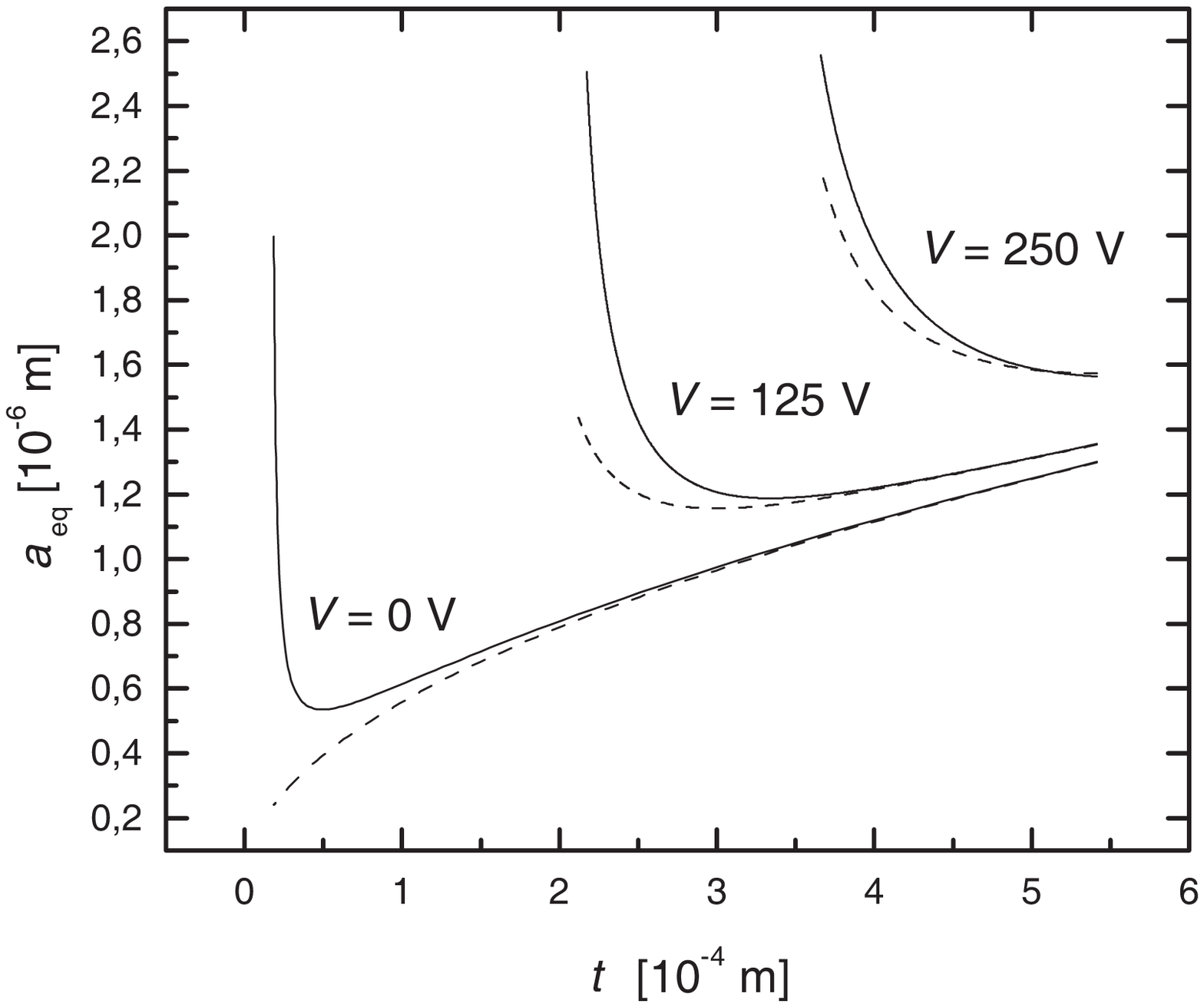}\vspace{3mm}
\caption{Exact numerical (full lines) and approximate (dashed
lines) results for $a_{\rm eq}(t)$ at different values of
potential difference $V$ and $B=0.02$.\label{artA1:fig:weqd}}
\end{minipage}
\end{center}
\end{figure}
If the system is short-circuited, the equilibrium domain pattern
is symmetric, i.e. $A_{\rm eq} = 0$. The shape factor $R_{\rm
eq}^0$  and from it also the value of $a_{\rm eq}^0$ can be found
by numerical methods. As an example, the full lines in
Fig.\,\ref{artA1:fig:weqd} shows the $a_{\rm eq}({\cal D})$
dependence at constant $h/d$ and different values of potential
difference $V$. As it was shown in the previous
paper\cite{ArtA1:Kopal}, the critical thickness $h_{\rm crit}$
can be defined by the formula
\[
    h_{\rm crit}
    \cong
    4\,\varepsilon_0\sigma_w\sqrt{\varepsilon_a\varepsilon_c}/P_0^2
\]
so that for $h\gg h_{\rm crit}$ the interaction energy of sample
surfaces can be neglected. Then the minimum energy occurs for
\begin{equation}
    a^0_{\rm eq} =
        \frac{1}{P_0}\left[\frac{\pi^3\varepsilon_0
            \left(\varepsilon_d + \sqrt{\varepsilon_a\, \varepsilon_c}\right)\,\sigma_w}{8 S_0
            (1+B)}\right]^{1/2}\,\sqrt{t} \ . \label{artA1:eq:w0}
\end{equation}
and the dependence $a_{\rm eq}^0(t)$  in this approximation is
shown by a dashed line with $V=0$\,V in the
Fig.\,\ref{artA1:fig:weqd}. The approximate results in the
Fig.\,\ref{artA1:fig:weqd} with $V=250\,$V and $V=500\,$V  are
based on Eqs.\,(\ref{artA1:eq:aeqv}) and (\ref{artA1:eq:reqv}). In
these numerical calculations we have used the following values
which roughly apply to crystals of $\rm RbH_2PO_4$ below the
transition temperature: $P_0 = 5.7\times 10^{-2}\,\rm C\,m^{-2}$,
$\varepsilon_a = 10, \varepsilon_c = 100$, $\varepsilon_d = 10$,
$h_{\rm crit} = 5.4\times 10^{-8}\,\rm m$. The value $\sigma_w =
5\times 10^{-3}\,\rm J\,m^{-2}$ is often considered typical for
ferroelectrics.

\section{Response of domain structute to external electric field}\label{artA1:sRtEF}

\begin{figure}[t]
\begin{center}
 \includegraphics[width=11cm]{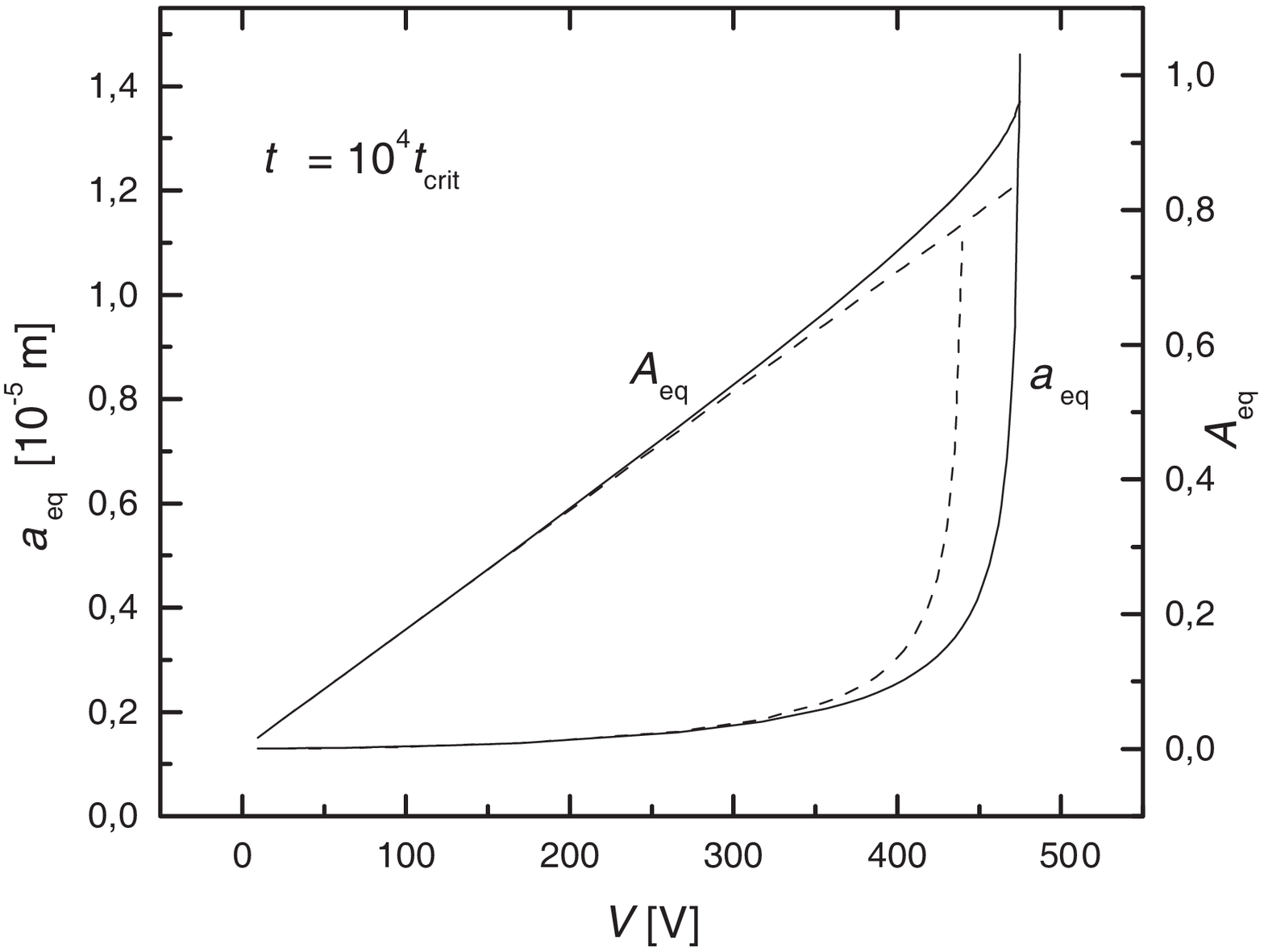}\vspace{3mm}
\begin{minipage}{11cm}
\caption{Exact numerical (full lines) and approximate (dashed
lines) results for $a_{\rm eq}(V)$ and $A_{\rm eq}(V)$ at $B=0.02$
and $t = 10^4 t_{\rm crit}$. \label{artA1:fig:weqv}}
\end{minipage}
\end{center}
\end{figure}
When an external potential difference $V$ is applied, the
asymmetry parameter becomes nonzero and at the same time the
period $a = (a_+ + a_-)/2$ changes. Both these quantities can be
found again by determining the minimum of $G$ given by
Eq.\,(\ref{artA1:eq:depolenergy}) numerically. Full lines in
Fig.\,\ref{artA1:fig:weqv} demonstrate both these dependencies for
the following numerical values: $B = 0.02$, $t =  10^4\,t_{\rm
crit}= 10^4\,(1 + B)\, h_{\rm crit}$. Starting from a certain
applied voltage the domain spacing $a$ grows very fast with
increasing $V$. To discuss the macroscopic properties of the
sample such as permittivity, the dependence $A_{\rm eq}(V)$ is
more important. We recognize that in a considerable region of the
applied voltage this dependence is almost linear.

The $A_{\rm eq}(V)$, $a_{\rm eq}(V)$ resp. $R_{\rm eq}(V)$
dependence can be approximated by an explicit formula if the
following inequalities are satisfied: $A\ll~1$, $BR\gg~1$ and
$Rc\gg~1$. Then it holds
\begin{equation}
    A_{\rm eq}(V) \cong
    \varepsilon_0 \varepsilon_d
    \left( 1 + B \right)
    \frac{
        V/t
    }{
        P_0
    }
    \left[
        B -
        \frac{
            2 \log 2
            \left(
                \varepsilon_d + B\,\varepsilon_c
            \right)
        }{
            R^0_{\rm eq}
            \left(
                \varepsilon_d + \sqrt{\varepsilon_a \varepsilon_c}
            \right)
        }
    \right]^{-1}
    \label{artA1:eq:aeqv}
\end{equation}
\begin{equation}
    R_{\rm eq}(V) \cong R_{\rm eq}^0 \sqrt{1 - \frac{\pi^2 \ln{2}}{4\,S_0}\,
    A^2_{\rm eq}(V)}\ . \label{artA1:eq:reqv}
\end{equation}
These approximations, shown in Fig.\,\ref{artA1:fig:weqv} by
dashed lines, are based on the limit of the sum in
Eq.\,(\ref{artA1:eq:depolenergy}).

\section{Extrinsic permittivity}\label{artA1:sEP}

The nonzero value of the asymmetry parameter means that an extra
bound charge is deposited on the electrodes due to the domain wall
shifts and this in turn represents an increase of effective
permittivity of the whole system crystal plus both surface layers.
The increase of the electrostatic energy when domain walls leave
their original equilibrium positions for $E\neq 0$ serves as the
source of a restoring force when field is again reduced to zero.
The calculations show that the effective permittivity
$\varepsilon_{\rm eff}$ defined by the total capacitance per unit
area $C = \varepsilon_0\varepsilon_{\rm eff}/t$ equals
\begin{equation}
    \varepsilon_{\rm eff} =
    \frac{
        \varepsilon_c
    \left(1 + B\right)
    }{
        1 + B\varepsilon_c/\varepsilon_d
    }
    +
    \frac{
        1 + B
    }{
        1 + B \varepsilon_c/\varepsilon_d
    }
    \left[
        \frac{B}{\varepsilon_d} -
        \frac{
            2 \ln 2
            \left(1 + B\varepsilon_c/\varepsilon_d\right)
        }{
            R^0_{\rm eq}\,
            \left(
                \varepsilon_d
                +
                \sqrt{\varepsilon_a\,
                \varepsilon_c}
            \right)
        }
    \right]
    ^{-1}
    \label{artA1:eq:eeff}
\end{equation}
In this formula the first term on the right-hand side represents the
intrinsic part of permittivity, given by linear dielectric response of
the sample and of the surface layers when domain walls are kept at rest.
The second term is the contribution of domain walls displacement to
effective permittivity, often referred to as extrinsic part of
permittivity.

\section{Discussion}

Numerous data are available on domain wall contributions to
permittivity in single crystals of ferroelectrics and also on
extrinsic contributions to piezoelectric coefficients in
ferroelectrics which are simultaneously ferroelastic. Our
calculations indicate that depolarizing energy can be an effective
source of restoring force whose existence is a condition for such
contributions. In fact since the model assumes a regular system of
planar domain walls, it is suitable in particular for ferroelastic
ferroelectrics such as crystals of the KDP family in which a dense
pattern of $180^{\circ}$ domains is known to
exist.\cite{ArtA1:Fouskova,ArtA1:Bornarel} It was found that in a wide
temperature range below the Curie point of crystals of $\rm
RbH_2PO_4$ and deuterated KDP, the piezoelectric coefficient
$d_{36}$ is greatly enhanced compared to its expected value for
single domain samples.\cite{ArtA1:Shuvalov} Recently, this was
confirmed by simultaneous measurements of permittivity
$\varepsilon_3$, elastic compliance $s_{66}$ and piezoelectric
coefficient $d_{36}$ of $\rm RbH_2PO_4$. A thorough discussion of
$d_{36}$ in this case will be the subject of a forthcoming paper
{Kopal \emph{et al.}}. Here we comment on the extrinsic part of
$\varepsilon_3$. For simplicity, let us assume that $\varepsilon_d
= \varepsilon_c$. This is not an unreasonable assumption since the
assumed surface layer for KDP-type samples can be supposed to have
a similar chemical composition as the bulk. Then the extrinsic
part of Eq.\,(\ref{artA1:eq:eeff}) reduces to
\[
    \left(
        \frac{B}{\varepsilon_c}
        -
        \frac{
            2\ln{2}\,(1+B)
        }{
            R_{\rm eq}^0
            \left(
                \varepsilon_c
                +
                \sqrt{\varepsilon_a\,\varepsilon_c}
            \right)
        }
    \right )^{-1}\ .
\]
Numerically, the second term in the brackets represents a small
correction to the first term when approximations
(\ref{artA1:eq:aeqv}) and (\ref{artA1:eq:reqv}) are valid. If it
is neglected, we obtain as an approximation
\begin{equation}
\varepsilon_{z,\rm extrinsic} \cong \frac{\varepsilon_c}{B} =
\varepsilon_c\,\frac{h}{d}\ . \label{artA1:eq:eextr}
\end{equation}
This shows that a very thin surface layer can lead to a
considerable extrinsic enhancement of permittivity. Nevertheless,
the simple implication: $d\to 0 \Rightarrow \varepsilon_{z,\rm
extrinsic}\to \infty$ is not correct, because the assumptions
needed for validity of (\ref{artA1:eq:aeqv}),
(\ref{artA1:eq:reqv}), (\ref{artA1:eq:eeff}) and
(\ref{artA1:eq:eextr}) are violated if $d$ is small enough.

A more general formulation of the restoring force can be used to
calculate the extrinsic part of $d_{36}$ for the same geometry of
domains. In a recent paper\cite{ArtA1:sidorkin} Sidorkin deduced
the dispersion law of wall contributions to permittivity, however,
in his treatment the existence of a surface layer is not
explicitly considered.

It was shown beyond any doubt that small motions of $90^{\circ}$
domain walls are responsible for a considerable enhancement of
permittivity $\varepsilon_3$ and piezoelectric coefficients
$d_{31}, d_{33}$ in poled ferroelectric
ceramics.\cite{ArtA1:Arlt1,ArtA1:Zhang} One of the sources of the
restoring force responsible for these wall contributions is the
elastic energy at grain boundaries.\cite{ArtA1:Arlt2} Since these
boundaries may differ in chemical composition from the bulk of
grains\cite{ArtA1:Heydrich}, surface layers can be expected to form
so that the mechanism proposed in the present paper may also play
a role in ceramic samples.


\end{document}